\documentclass{iopjournal}

\begin{document}

\articletype{Perspective} 

\title{Quantum Optimization Algorithms for Strongly Correlated Many-Body Systems}

\author{
G. E. L. Pexe$^{1*}$,
L. A. M. Rattighieri$^{2}$,
P. M. Prado$^{3}$,
A. R. Fritsch$^{1}$ and
F. F. Fanchini$^{3,4}$
}

\affil{$^1$Instituto de F\'{\i}sica de S\~ao Carlos, Universidade de S\~ao Paulo, CP 369, 13560-970 S\~ao Carlos, Brazil}

\affil{$^2$Instituto de F\'{\i}sica Gleb Wataghin, Universidade Estadual de Campinas, 13083-859 Campinas, SP, Brazil}

\affil{$^3$Faculty of Sciences, UNESP - S{\~a}o Paulo State University, 17033-360 Bauru-SP, Brazil}

\affil{$^4$QuaTI - Quantum Technology \& Information, 13560-161 S\~ao Carlos-SP, Brazil}

\affil{$^*$Author to whom any correspondence should be addressed.}

\email{guilherme.pexe@ifsc.usp.br}


\begin{abstract}
This perspective article analyzes the potential and critical challenges of employing quantum optimization algorithms to investigate phase transitions in quantum many-body systems during the Noisy Intermediate-Scale Quantum era. The simulation of strongly correlated systems is frequently intractable on classical computers due to the exponential growth of the Hilbert space and the fermionic sign problem. In this context, we review and compare the performance of traditional Variational Quantum Algorithms, such as the Variational Quantum Eigensolver and the Quantum Approximate Optimization Algorithm, against emerging heuristic approaches, specifically Feedback-based Quantum Algorithms, such as FALQON. We explore the applicability of these methods in the study of open phenomena in condensed matter physics, including Deconfined Quantum Criticality, strange metals, Many-Body Localization, topological phase transitions, and quantum spin liquids. We discuss how fundamental operational bottlenecks, notably expressibility- and noise-induced barren plateaus, severely compromise gradient-based optimization. We conclude that deterministic feedback-guided methods provide geometrically more robust trajectories for navigating the energy landscape of these systems, arguing that further advancement in the field will rely on deep hybridization and physics-informed circuit co-design towards fault tolerance.
\end{abstract}

\section{Introduction}

The accurate description of quantum many-body systems remains one of the central and most complex challenges in modern physics. In regimes where interactions between constituents—whether electrons, spins, or cold atoms—dominate over kinetic energy or thermal fluctuations, exotic collective behaviors emerge that cannot be explained by mean-field theories or standard perturbative methods. The fundamental difficulty in simulating these strongly correlated systems lies in the mathematical structure of quantum mechanics: the dimension of the Hilbert space grows exponentially with the number of degrees of freedom ($N$). For a system of spins-$1/2$, the dimension is $2^N$, making the exact storage of the wavefunction impossible for classical computers, even for modest systems of $N \approx 50$ particles. This exponential barrier, often referred to as the dimensionality catastrophe, precludes exact analytical treatments and severely limits numerical exact diagonalization \cite{Feynman1982, Lloyd1996}.

The scalability of simulation methods is of critical importance, particularly in the study of critical phenomena and quantum phase transitions. The universality and critical exponents characterizing these transitions only fully manifest in the thermodynamic limit ($N \to \infty$). Consequently, the inability to simulate large lattices restricts the analysis of long-range correlations and the precise extraction of macroscopic properties through Finite Size Scaling. Classical numerical approaches, while powerful, face insurmountable obstacles in specific regimes. The Quantum Monte Carlo (QMC) method, for instance, suffers from the infamous sign problem when dealing with fermions or frustrated spin systems, where the probability weight function becomes negative or complex, causing statistical precision to degrade exponentially \cite{Troyer2005}. Alternatively, methods based on Tensor Networks, such as the Density Matrix Renormalization Group (DMRG), are extremely efficient for 1D systems but struggle to capture the growth of entanglement in higher dimensions, where the entanglement entropy scales with the subsystem volume rather than its area \cite{Schollwock2011, Orus2014}.

Against this backdrop of classical limitations, quantum computing emerges not merely as an alternative but as the natural platform for simulating nature, as envisioned by Feynman. Currently, we operate in the Noisy Intermediate-Scale Quantum (NISQ) era, characterized by processors featuring dozens to hundreds of qubits but lacking full quantum error correction \cite{Preskill2018}. The noise inherent in logic gates and the limited coherence time of qubits impose severe restrictions on executable circuit depth. This has necessitated a paradigm shift toward Variational Quantum Algorithms, such as the Variational Quantum Eigensolver (VQE) and the Quantum Approximate Optimization Algorithm (QAOA). These hybrid approaches utilize the quantum computer for the costly task of preparing complex states and estimating expectation values, while delegating the optimization of circuit parameters to a classical processor, attempting to mitigate noise effects through shallower circuits \cite{Peruzzo2014, Farhi2014, Cerezo2021}.

However, the straightforward application of VQE and QAOA is not a panacea. Classical optimization within the energy landscapes of correlated systems faces formidable challenges, notably the phenomenon of barren plateaus, where cost function gradients vanish exponentially with system size, rendering training infeasible. The primary objective of this perspective article is to outline the main open problems involving phase transitions in quantum many-body systems and discuss how quantum optimization algorithms can provide new avenues for studying such phenomena. Moving beyond the standard implementation of these algorithms, we discuss how novel approaches, specifically Feedback-based Quantum Algorithms (FQAs), may overcome the limitations of gradient-based optimizers. Unlike static variational methods, FQAs—inspired by quantum control and system dynamics—offer alternative routes to navigate complex energy landscapes, enabling the investigation of ground and excited states in complex quantum materials \cite{Magann2021}.

\section{Quantum optimization algorithms in a NISQ perspective}

To address the complex open problems involving phase transitions in many-body systems, it is essential first to establish the algorithmic tools at our disposal. The quest for quantum advantage in noisy intermediate-scale devices has been predominantly driven by variational algorithms and control-based methods. In this section, we discuss three central paradigms: VQE, QAOA, and the emerging FQAs. We analyze their viability in the face of hardware constraints, promising variants, and, fundamentally, how their mechanisms can be adapted to probe condensed matter physics, paving the way for the study of critical phenomena.

\begin{figure}[t]
    \centering
    \includegraphics[width=0.6\linewidth]{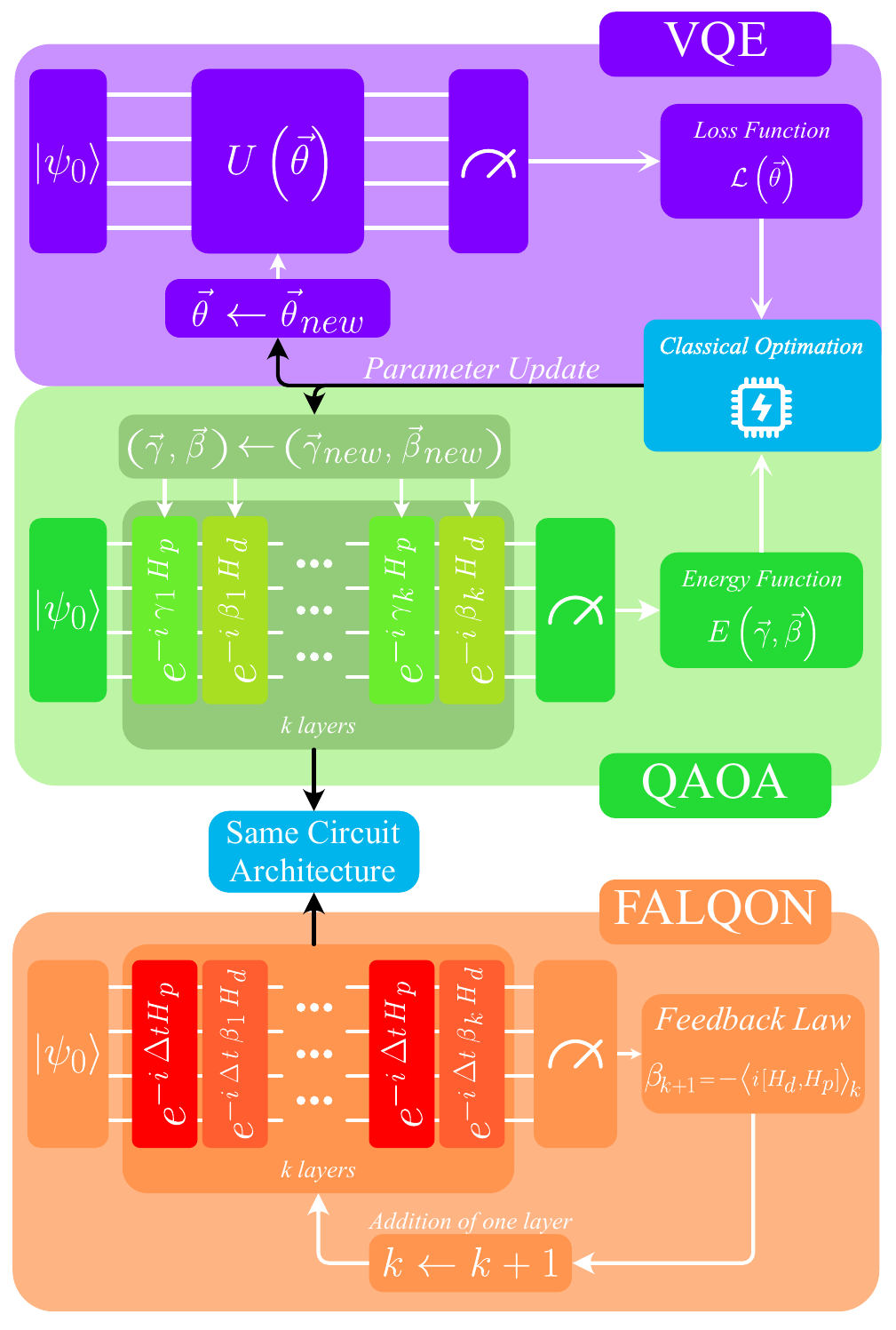}
    \caption{Comparative schematic of the main quantum optimization algorithms from a NISQ perspective. (a) In VQE, a variational circuit prepares an ansatz and measures the expectation value of the Hamiltonian, while a classical optimizer iteratively updates the parameters. (b) In QAOA, an alternating sequence of evolutions under the problem Hamiltonian $H_p$ and the mixer Hamiltonian $H_d$ is applied for a fixed depth $p$, followed by classical optimization of the angles. (c) In FALQON, the dynamics are guided by feedback based on measuring the commutator $i[H_d,H_p]$, eliminating the external classical optimizer and introducing an iterative evolution with progressive growth of the circuit depth.}
    \label{fig:nisq_algorithms}
\end{figure}

Figure~\ref{fig:nisq_algorithms} summarizes the structural differences among three central classes of quantum optimization algorithms in the NISQ regime. Although VQE and QAOA share the hybrid variational philosophy, both rely on a classical optimization loop to tune parameters, rendering them sensitive to sampling costs and measurement errors. In contrast, feedback-based algorithms, such as FALQON, replace the external optimizer with a feedback rule derived from the system's own observable quantities.

\subsection{Variational Quantum Eigensolver}

VQE remains the cornerstone of hybrid variational algorithms. Its operation relies on the Rayleigh-Ritz variational principle, where a quantum processor prepares a trial state $\ket{\psi(\vec{\theta})}$ parameterized by an ansatz, and a classical coprocessor optimizes the parameters $\vec{\theta}$ to minimize the energy expectation value $E = \bra{\psi(\vec{\theta})} H \ket{\psi(\vec{\theta})}$. The primary advantage of VQE lies in its flexibility and low circuit depth compared to Quantum Phase Estimation. Furthermore, it possesses inherent resilience against certain systematic errors, such as imperfect gate calibration, since the classical optimizer can adjust parameters to mitigate these flaws. Among the variants developed to circumvent expressibility limitations, notable examples include ADAPT-VQE \cite{Tang_2021}, which dynamically constructs the ansatz to save circuit depth, and Subspace VQE (SS-VQE), focused on determining excited states \cite{Grimsley2019_NatComm_ADAPT, Nakanishi2019_SSVQE}.

However, VQE performance on NISQ hardware is severely impacted by sampling noise and readout error. The need to estimate the Hamiltonian expectation value requires a massive number of measurements that scales inversely with the square of the target accuracy. This sampling cost, combined with the phenomenon of noise-induced barren plateaus, renders the cost landscape flat and often indistinguishable from background noise, precluding the precise convergence required to detect subtle energy gaps \cite{Peruzzo2014, McClean2018_Barren, Cerezo2021, Tilly2022}.

\subsection{Quantum Approximate Optimization Algorithm}

Originally conceived for combinatorial optimization, QAOA is interpreted in physics as a digitized Trotterized adiabatic evolution. The algorithm alternately applies a problem Hamiltonian $H_p$ and a mixer Hamiltonian $H_d$ for $p$ layers, providing theoretical guarantees of convergence in the infinite-depth limit. The primary advantage of QAOA is its physical interpretability, where the circuit directly reflects the problem's connectivity structure. Furthermore, variants such as Recursive QAOA (R-QAOA) \cite{Bravyi2020_RQAOA} eliminate variables to reduce graph complexity, while Multi-angle QAOA \cite{Herrman2022_maQAOA} increases expressibility by allowing distinct parameters for each Hamiltonian term.

The most impactful hardware flaw for QAOA in the NISQ regime is the limited fidelity of two-qubit gates (such as CNOT or CZ) coupled with the restricted topological connectivity of the chips. Unlike VQE, where the ansatz can be hardware-tailored, rigid QAOA requires the exact implementation of $H_p$. If the processor's connectivity does not match the problem's interaction graph, the necessary insertion of SWAP networks drastically increases circuit depth. Consequently, increasing the number of layers $p$ to improve the theoretical approximation frequently degrades the practical outcome due to the accumulation of gate errors, which destroys constructive interference \cite{Farhi2014, Wurtz2020}.

\subsection{Feedback-based Quantum Algorithms}

Feedback-based Quantum Algorithms, exemplified by the FALQON (Feedback-based ALgorithm for Quantum OptimizatioN) framework, represent a paradigm shift by removing the external classical optimizer in favor of a deterministic update rule based on measuring the commutator $\ev{i[H_B, H_C]}$ \cite{PhysRevB.110.224422}. This approach, inspired by Lyapunov control theory, eliminates the complex landscape of stochastic multidimensional optimization. Two variants are particularly relevant for improving the performance of these algorithms: Time-Rescaled FALQON (TR-FALQON) \cite{qc91-5mj2} and Imaginary-Time-Enhanced FALQON (ITE-FALQON) \cite{Long2025_ITEFALQON}. TR-FALQON introduces a dynamic rescaling of the evolution time based on the magnitude of the energy derivative, solving issues of oscillation and slow convergence in regions with a dense energy spectrum. ITE-FALQON, on the other hand, incorporates concepts of imaginary-time evolution by modifying the feedback rule to mimic non-unitary cooling dynamics, proving effective in avoiding local minima traps that plague purely unitary approaches \cite{Long2025_ITEFALQON}.

The primary challenge for FQAs on NISQ devices lies in their sensitivity to measurement errors and the latency of the feedback loop. Because the update for the subsequent step depends deterministically on the precise measurement of the previous step, unmitigated readout errors can steer the algorithm along a trajectory that increases the system's energy. Additionally, the circuit depth grows progressively as new interactions are performed, due to the successive addition of layers. This structural increase implies a greater accumulation of gate errors and greater exposure to decoherence, exacerbating the limitations imposed by the finite coherence time of the qubits. If the coherence time is short compared to the latency required to process the classical feedback and reconfigure the circuit, quantum information degrades between application layers. Despite these challenges, FQAs show promise in preparing specific excited states without the orthogonalization cost required by VQE.

\section{Applications in Condensed Matter Physics and Quantum Phase Transitions}

Having established the theoretical architectures and operational challenges of VQE, QAOA, and FQAs in the NISQ regime, this section fulfills the central objective of this perspective: to examine the practical applicability of these tools to frontier problems. Moving beyond the mitigation of the intrinsic dimensional bottleneck of classical exact diagonalization or Quantum Monte Carlo methods, we investigate how the unique mechanisms of these algorithms—such as the adaptive construction of ansätze or feedback-guided dynamic evolution—offer robust probes for emergent properties in strongly correlated systems.

\begin{figure*}[htbp]
    \centering
    \includegraphics[width=0.89\textwidth]{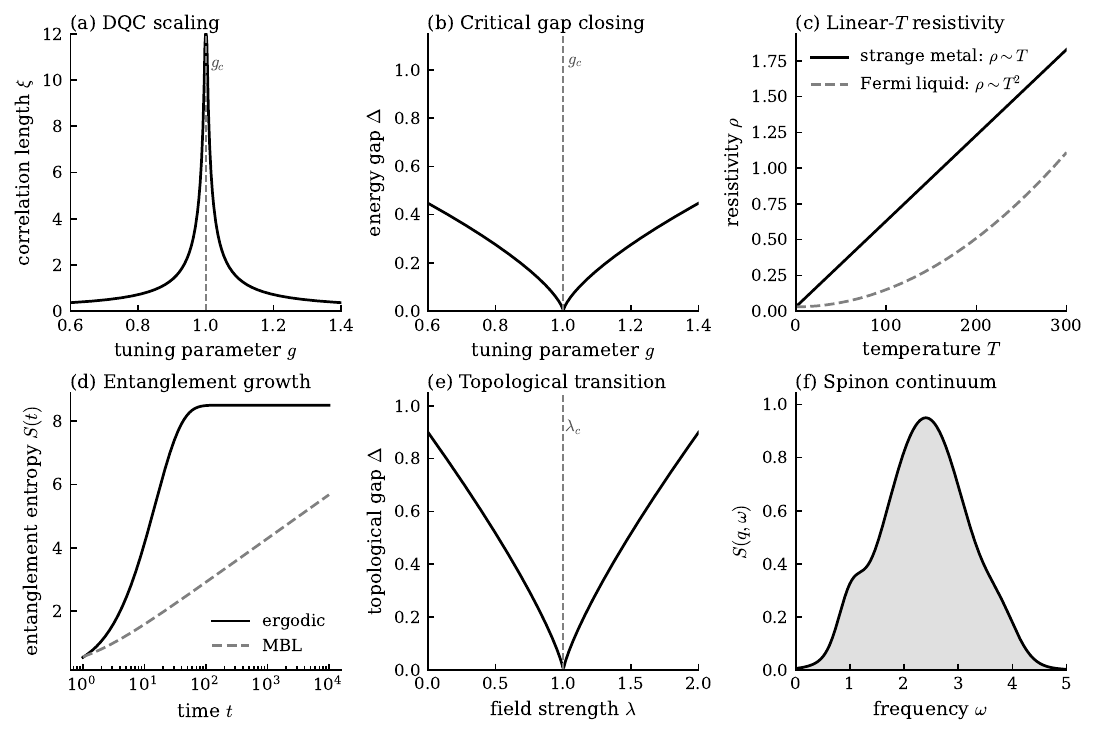}
    \caption{Phenomenological signatures of quantum phase transitions and strongly correlated states targeted for investigation via quantum algorithms in the NISQ regime. (a) Divergence of the correlation length $\xi$ and (b) algebraic closing of the energy gap $\Delta$ in the vicinity of a Deconfined Quantum Criticality (DQC) point $g_c$. (c) Anomalous transport in the strange metal phase, evidenced by $T$-linear resistivity ($\rho \propto T$), in contrast to the quadratic dependence of a Fermi liquid. (d) Non-equilibrium dynamics of the entanglement entropy $S(t)$ following a global quench: the Many-Body Localization (MBL) phase exhibits ultra-slow logarithmic growth, while the ergodic phase rapidly saturates following a volume law. (e) Closing of the topological gap $\Delta$ as a function of a transverse magnetic field $\lambda$, marking anyon condensation at $\lambda_c$. (f) Dynamic structure factor $S(\mathbf{q}, \omega)$ illustrating the broad scattering continuum characteristic of spinon fractionalization in a Quantum Spin Liquid.}
    \label{fig:phenomena_panel}
\end{figure*}

In this context, we outline below five fundamental open problems in many-body physics, whose Quantum Phase Transitions at zero temperature ($T=0$) frequently defy the traditional Landau-Ginzburg-Wilson paradigm (their phenomenological signatures are summarized in Figure~\ref{fig:phenomena_panel}). We discuss the physics underlying each phenomenon and highlight how the flexibility of quantum optimization algorithms offers novel routes to investigate states with long-range entanglement, from disorder-induced localization to the exotic nature of quantum spin liquids.

\subsection{Deconfined Quantum Criticality}

The traditional Landau-Ginzburg-Wilson (LGW) paradigm establishes that continuous phase transitions are characterized by the fluctuations of a local order parameter associated with spontaneous symmetry breaking. Within this framework, direct transitions between two phases that break independent symmetries—such as an antiferromagnetic Néel state (breaking $SU(2)$ spin rotation symmetry) and a Valence Bond Solid (VBS) state (breaking the discrete lattice translation and rotation symmetries)—should generically occur via a first-order transition or through an intermediate coexistence phase. However, the DQC theory \cite{Senthil2004_Science_DQC} proposes an exotic scenario where a genuinely continuous transition can occur between these phases without the need for parameter fine-tuning.

At this non-LGW critical point, the relevant fluctuations are not described by the original order parameters but rather by fractionalized degrees of freedom (bosonic spinons) strongly coupled to an emergent $U(1)$ gauge field. The transition is governed by the suppression of topological defects (hedgehogs or monopoles) in the Néel order parameter that carry the VBS charge, a non-trivial consequence of the many-body Berry phases. The prototypical microscopic model introduced to investigate this phenomenon is the $J$-$Q$ model defined on a two-dimensional square lattice \cite{Sandvik2007_PRL_JQ}:
\begin{equation}
    H_{J-Q} = -J \sum_{\langle i,j \rangle} P_{ij} - Q \sum_{\langle i,j \rangle \langle k,l \rangle \in \square} P_{ij} P_{kl}
\end{equation}
where $P_{ij} = \frac{1}{4} - \mathbf{S}_i \cdot \mathbf{S}_j$ is the spin-$1/2$ singlet projection operator on the bond $\langle i,j \rangle$, and the $Q$ term typically couples parallel bonds on adjacent plaquettes.

The central open problem in this domain is to determine the exact nature of the critical point $g_c = (J/Q)_c$ in the thermodynamic limit. Large-scale Quantum Monte Carlo simulations exhibit remarkable finite-size scaling violations and drifting critical exponents (drifting scaling dimensions) \cite{Shao2016_Science_Pseudo}. This has sparked intense debate over whether the transition is described by a genuine continuous Conformal Field Theory (CFT)—possibly a non-unitary complex CFT—or if it is a regime of "pseudo-criticality," consisting of a weakly first-order transition that only reveals itself at massively large length scales (beyond current numerical reach) due to the slow, dangerously irrelevant nature of the monopole operators.

The simulation of DQC on quantum computers presents both formidable challenges and opportunities to bypass classical limitations. To explore the DQC critical point, VQE can, in principle, be employed to prepare the ground state $|\psi(\vec{\theta})\rangle$ by sweeping the $J/Q$ ratio. However, as the system approaches the DQC point, the spectral energy gap closes algebraically ($\Delta \propto L^{-z}$) and the correlation length diverges ($\xi \to \infty$)—dynamics illustrated in Figures \ref{fig:phenomena_panel}(a) and \ref{fig:phenomena_panel}(b). In such critically entangled regimes, traditional physics-agnostic variational ansätze (Hardware-Efficient Ansätze - HEA) suffer severely from entanglement-induced barren plateaus, rendering gradient optimization exponentially difficult.

In this scenario, pure feedback-based approaches may face geometric stagnation: if the FALQON mixer Hamiltonian fails to break specific symmetries of the initial subspace, the commutator vanishes prematurely, preventing the full exploration of the Hilbert space down to the ground state. To circumvent this, ITE-FALQON emerges as the structurally most robust strategy \cite{Long2025_ITEFALQON}. By incorporating imaginary-time dynamics into the update rule, the algorithm mimics a non-unitary cooling that is less susceptible to being trapped in symmetry-protected local minima. The growth of the ansatz is guided adaptively and deterministically, making ITE-FALQON particularly well-suited to capture the intricate long-range correlations and spinon deconfinement near the DQC critical zone, bypassing the optimization bottleneck that paralyzes standard variational methods.

\subsection{Strange Metal Phases and Strongly Correlated Superconductivity}

The strange metal phase emerges ubiquitously in the phase diagram of strongly correlated materials, typically fanning out above a Quantum Critical Point at finite temperatures, as observed in superconducting cuprates and heavy fermions. In visceral contrast to Landau's Fermi liquid theory, where low-energy physics is governed by coherent, long-lived quasiparticles ($\tau \propto T^{-2}$), the strange metal exhibits a total absence of well-defined quasiparticles. Its most striking phenomenological signature is $T$-linear resistivity ($\rho \propto T$), in sharp opposition to the quadratic behavior of a Fermi liquid [Fig. \ref{fig:phenomena_panel}(c)]. This is a direct consequence of a universal inelastic scattering rate that saturates at the so-called Planckian limit, $\tau \sim \hbar / (k_B T)$, where dissipation is governed exclusively by fundamental constants and temperature.

In this ultra-correlated fluid regime, it is postulated that the same critical fluctuations (spin or charge) responsible for destroying quasiparticle coherence act as the mediating glue for unconventional superconducting pairing (such as $d$-wave symmetry pairing in cuprates). As an analytically tractable limit for many-body systems without quasiparticles, the Sachdev-Ye-Kitaev (SYK) model has become a fundamental theoretical paradigm \cite{Sachdev1993_PRL_SYK}:
\begin{equation}
    H_{SYK} = \frac{1}{(2N)^{3/2}} \sum_{i,j,k,l=1}^{N} J_{ijkl} \chi_i \chi_j \chi_k \chi_l
\end{equation}
where $\chi_i$ represent Majorana fermions (satisfying $\{\chi_i, \chi_j\} = \delta_{ij}$) and $J_{ijkl}$ are infinite-range random tensor couplings drawn from a Gaussian distribution. In the large-$N$ thermodynamic limit, the SYK model is exactly solvable and exhibits emergent conformal symmetry at low energies. Furthermore, it saturates the upper bound on the quantum chaos rate (thermal Lyapunov exponent $\lambda_L = 2\pi k_B T / \hbar$), connecting the Planckian dissipation of strange metals to black hole dynamics via holographic duality \cite{altland2026quantumchaosholographicprinciple}.

The classical simulation of fermionic thermalization dynamics is rendered intractable by the exponential growth of entanglement. On quantum hardware, the SYK Hamiltonian can be efficiently mapped to qubit registers using Jordan-Wigner or Bravyi-Kitaev transformations. To circumvent the unfeasible depth of Trotter-Suzuki circuits in the study of non-equilibrium dynamics, variational algorithms such as QAOA adapted for time evolution (Variational Quantum Time Evolution - VQTE) offer a promising alternative \cite{Yuan2019_Quantum_VQTE}. 

With shallow parameterized circuits, it is possible to simulate time evolution following a quench and investigate the fast scrambling of quantum information, a defining characteristic of the SYK model. The central physical metric to be extracted from the quantum computer is the Out-of-Time-Ordered Correlator (OTOC), defined as $C(t) = \langle \psi | W^\dagger(t) V^\dagger(0) W(t) V(0) | \psi \rangle$. The exponential decay of OTOCs on quantum platforms can provide direct evidence of the Kitaev chaos bound \cite{Maldacena2016_JHEP_ChaosBound} and illuminate the intricate transition between non-Fermi liquid fluctuations and correlated superconducting condensation.

\subsection{Many-Body Localization Transition}

Strongly interacting quantum many-body systems, when subjected to a sufficiently strong disordered potential, can fail completely to act as their own thermal baths. This phenomenon, known as Many-Body Localization, represents a robust breakdown of conventional statistical mechanics and a direct violation of the Eigenstate Thermalization Hypothesis \cite{Abanin2019_RMP_MBL}. Unlike Anderson localization for non-interacting particles, the MBL phase preserves interactions, yet the system acquires a macroscopic set of emergent Local Integrals of Motion (l-bits or LIOMs). These constants of motion prevent charge and energy transport, causing the system to retain the memory of its initial conditions at infinite times.

The canonical model for investigating this physics is the one-dimensional spin-$1/2$ Heisenberg chain (XXX model) in the presence of a random magnetic field:
\begin{equation}
    H = J \sum_{i} \vec{\sigma}_i \cdot \vec{\sigma}_{i+1} + \sum_{i} h_i \sigma_i^z
\end{equation}
where $\vec{\sigma}_i = (\sigma_i^x, \sigma_i^y, \sigma_i^z)$ are the Pauli matrices and the local field amplitudes $h_i$ are sampled from a uniform distribution $[-W, W]$, with $W$ being the disorder strength. The spectrum exhibits a many-body mobility edge, separating high-energy ergodic states from localized ones. 

The nature of the disorder-driven transition at $W_c$ between the thermal phase (whose eigenstates exhibit volume-law entanglement entropy, $S \propto L$) and the MBL phase (whose eigenstates strictly obey an area law, $S \propto \text{constant}$) is one of the most acute and heavily debated problems in contemporary physics. Exact simulations suffer from severe finite-size effects, and recently, theories based on "thermal avalanches" (where rare ergodic inclusions destabilize the entire system) have raised the fundamental question of whether the MBL phase is genuinely stable in the thermodynamic limit for dimensions $d > 1$, or even in $1$D, suggesting that what is observed might merely be an ultra-long-lived prethermal regime \cite{DeRoeck2017_PRB_Avalanches}.

The computational probing of the MBL phase requires access to highly excited mid-spectrum eigenstates, which poses a formidable challenge for classical methods such as DMRG. On quantum hardware, variational \textit{ansätze} such as Variational Quantum Deflation \cite{Higgott2019_Quantum_VQD} can be employed iteratively to extract excited states by optimizing a modified cost function: $C_k(\vec{\theta}) = \langle \psi(\vec{\theta}) | H | \psi(\vec{\theta}) \rangle + \sum_{j<k} \beta_j |\langle \psi(\vec{\theta}) | \phi_j \rangle|^2$, where $\beta_j$ are penalty terms that enforce orthogonality with the previously found states. 

Furthermore, the topology of Parameterized Quantum Circuits, such as the QAOA, serves as a fundamental dynamical probe for the transition: in the ergodic phase, the entangling gate depth $p$ required to faithfully prepare an excited eigenstate grows exponentially with the system size due to the volume law. In stark contrast, in the MBL phase (area law), low-depth circuits ($p \sim \mathcal{O}(1)$) are expressive enough to approximate the localized wavefunctions. Additionally, quantum computers can directly simulate the time evolution of product states (global quenches), allowing the measurement of the unambiguous signature of MBL: the ultra-slow logarithmic growth of the entanglement entropy, $S(t) \propto \ln(t)$, generated by the dephasing of residual interactions between the l-bits, which sharply contrasts with the rapid volume-law-governed saturation of the ergodic phase [Fig. \ref{fig:phenomena_panel}(d)].

\subsection{Topological Phase Transitions and Global Entanglement}

Phases with intrinsic topological order, whose central paradigm is Kitaev's Toric Code \cite{Kitaev2003_AnnPhys_ToricCode}, entirely evade the Landau-Ginzburg-Wilson description as they lack a local order parameter. Instead, they are characterized by a ground-state degeneracy that depends on the topology of the underlying spatial manifold, long-range quantum entanglement, and quasiparticle excitations with fractional statistics (anyons). The stability of such a phase is diagnosed by a non-zero Topological Entanglement Entropy (TEE, $\gamma = \ln 2$), which acts as a universal sub-leading correction to the area law ($S = \alpha L - \gamma$) \cite{Kitaev2006_PRL_TEE}.

The canonical microscopic model subjected to an external magnetic field is given by:
\begin{equation}
    H = -J_e \sum_{s} A_s - J_m \sum_{p} B_p - \lambda \sum_{i} \sigma_i^x
\end{equation}
where the spins reside on the edges of a two-dimensional lattice, $A_s = \prod_{i \in s} \sigma_i^x$ is the vertex (star) operator measuring the electric charge $e$, and $B_p = \prod_{i \in p} \sigma_i^z$ is the plaquette operator measuring the magnetic flux $m$. In the topological limit ($\lambda = 0$), these operators commute, defining a robust energy gap that protects the defect-free ground state ($A_s = B_p = +1$). 

The introduction of a transverse magnetic field ($\lambda > 0$) introduces non-commutativity and provides kinetic energy to the anyonic $e$ excitations. When the field reaches a critical value $\lambda_c$, a quantum phase transition occurs, driven by anyon condensation \cite{Trebst2007_PRL_ToricPhase}. The topological gap closes, as illustrated in Fig.~\ref{fig:phenomena_panel}(e), anyons proliferate macroscopically, and the system collapses into a trivial polarized phase. Understanding the critical exponents of this transition (which maps to the 3D Ising universality class) and the non-local restructuring of multipartite entanglement requires analytical and numerical tools that go far beyond the constraints of conventional local field theories.

Simulating the dynamics and state preparation at topological phase boundaries faces a severe obstacle: the protection of gauge symmetry. In a quantum computer, operational errors or generic (hardware-efficient) variational ansätze quickly leak the state into non-physical sectors of the massive $2^N$-dimensional Hilbert space, violating the system's $\mathbb{Z}_2$ Gauss's law. 

In this context, the FALQON algorithm emerges as an intrinsically superior strategy. Being constructed via iterative dynamical evolution generated by commutators, the FALQON ansatz strictly inherits the symmetries shared by the target Hamiltonian and the initial state. If a driving Hamiltonian and an initial state are chosen to respect the gauge theory stabilizers (e.g., commuting with $B_p$ operators), the evolution of the FALQON circuit guarantees, by theoretical construction, unconditional permanence within the physical subspace. This facilitates the system's extrapolation across the topological phase boundary, tracking the gap closure and anyon condensation without corrupting global topological constraints.

\subsection{Transitions involving Quantum Spin Liquids}

Quantum Spin Liquids (QSLs) represent an exotic state of matter where intense geometric frustration and strong quantum fluctuations prevent any long-range magnetic ordering or translational symmetry breaking \cite{Savary2016_RepProgPhys_QSL}, even in the absolute zero temperature limit ($T=0$). The paradigmatic model for investigating this physics is the antiferromagnetic Heisenberg model on the Kagome lattice:
\begin{equation}
    H = J \sum_{\langle i,j \rangle} \vec{S}_i \cdot \vec{S}_j
\end{equation}
Classically, this system possesses a macroscopic ground-state degeneracy. Quantally, the state resolves into a highly entangled spin liquid, where the original magnetic degrees of freedom (local spins) fractionalize into neutral spin-$1/2$ quasiparticles (spinons), which interact strongly through emergent artificial gauge fields (often with $U(1)$ or $\mathbb{Z}_2$ symmetry).

The debate regarding the exact nature of the Kagome model's ground state—specifically whether it is a gapless $U(1)$ Dirac Quantum Spin Liquid (QSL) or a gapped $\mathbb{Z}_2$ topological QSL—remains one of the most challenging problems in many-body theory. Quantum phase transitions occur when microscopic perturbations (such as next-nearest-neighbor interactions $J_2, J_3$, or Dzyaloshinskii-Moriya interactions) destabilize the QSL in favor of conventional magnetic orders (such as $\mathbf{q}=0$ or $\sqrt{3} \times \sqrt{3}$ coplanar order) or VBS. Such transitions cannot be described by the Landau paradigm; they involve the condensation of fractionalized matter (the Higgs effect of spinons) or the proliferation of monopoles in the emergent gauge field, leading to the confinement of excitations.

The unambiguous identification of a topological ($\mathbb{Z}_2$) QSL phase in numerical simulations requires the calculation of the Topological Entanglement Entropy ($\gamma$), extracted from the scaling of the spatial entanglement entropy ($S_A = c L_A - \gamma$). On quantum hardware, after preparing the ground state via VQE, the reduced density matrix of the subsystem $\rho_A$ is exponentially large, rendering traditional state tomography unfeasible. However, the integration of VQE with advanced Classical Shadow Tomography protocols based on randomized measurements \cite{Huang2020_NatPhys_Shadows} allows for the efficient estimation of purity and second-order Rényi entropy ($S_2 = -\ln \text{Tr}[\rho_A^2]$), making the extraction of $\gamma$ viable.

Additionally, the experimental verification of QSLs in real materials (via inelastic neutron scattering) relies on detecting a scattering continuum of excitations, reflecting the fractionalized nature of spinon pairs, in contrast to the sharp magnon peaks in ordered systems. Hybrid variational algorithms and dynamical analogs based on QAOA can be adapted to simulate quantum spectroscopy on the Kagome lattice. By calculating time-domain correlation functions and applying the Fourier transform to obtain the dynamic structure factor $S(\mathbf{q}, \omega)$, as illustrated in Fig.~\ref{fig:phenomena_panel}(f), quantum simulators promise to directly map the spinon continuum \cite{Chiesa2019_NatPhys_Dynamics} and track the exact spectral signature of the spin-liquid phase destabilization through the quantum critical point.

\section{Critical Challenges}

\begin{figure}[t]
    \centering
    \includegraphics[width=\textwidth]{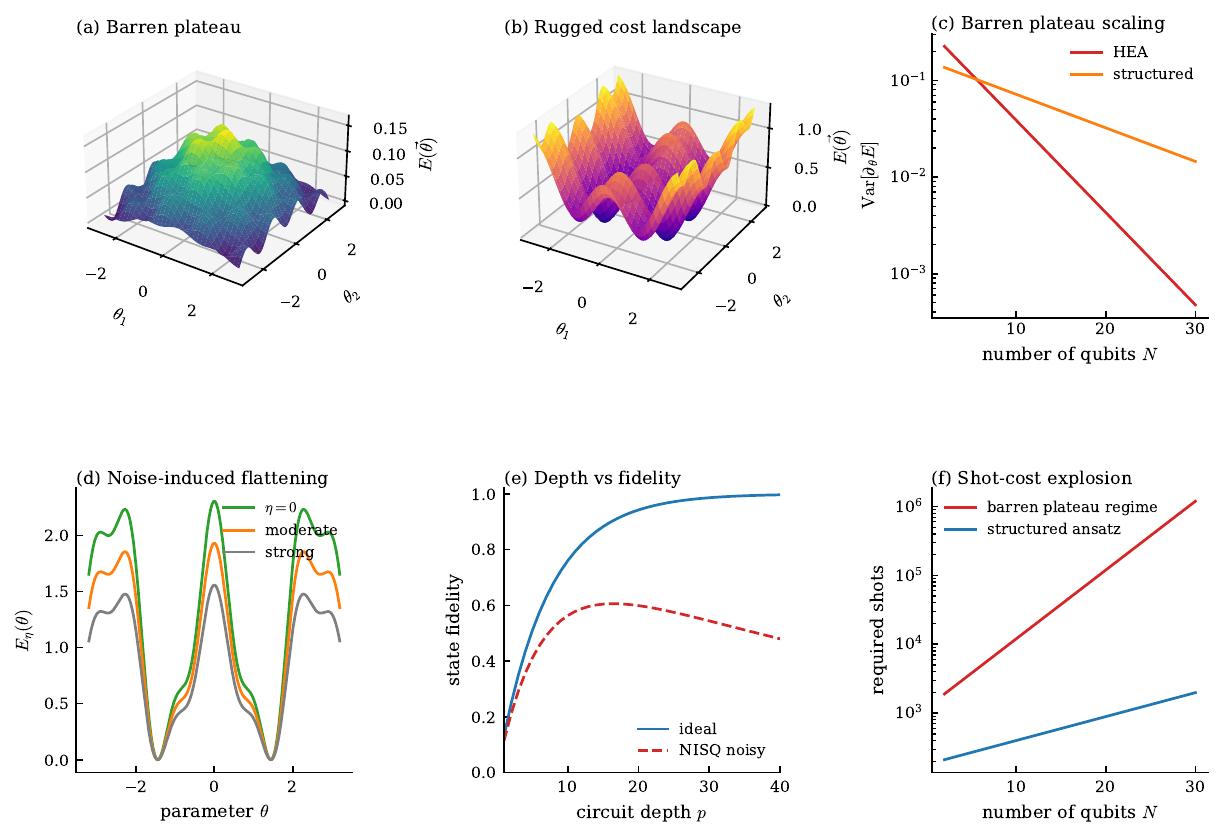}
    \caption{Representation of the fundamental challenges for variational quantum algorithms in condensed matter within the NISQ regime. 
    (a) Visualization of a barren plateau, characterized by an exponentially flat cost landscape that renders gradient-based methods unfeasible. 
    (b) Rugged cost landscape exhibiting multiple spurious local minima that trap classical optimizers. 
    (c) Scaling of the gradient variance $\text{Var}[\partial_\theta E]$ with the number of qubits $N$, demonstrating the severe exponential decay for Hardware Efficient Ansätze (HEA) compared to structured circuits. 
    (d) Effect of experimental noise ($\eta$) on the flattening of the energy landscape and the displacement of the global minimum. 
    (e) Trade-off between circuit depth and state fidelity: the presence of noise imposes a maximum useful depth limit before the dominance of decoherence. 
    (f) Explosion of the sampling cost (shots) required to resolve the energy surface in barren plateau regimes.}
    \label{fig:desafios}
\end{figure}

Although the applications outlined in the previous section highlight the transformative potential of variational and adaptive algorithms for investigating many-body physics, the practical implementation of these routines on NISQ-era hardware faces fundamental obstacles. The transition from idealized theory to the laboratory reveals severe limitations regarding the optimizability and noise resilience of parameterized circuits, as illustrated in the general panel of Fig.~\ref{fig:desafios}—challenges that must be overcome to achieve quantum advantage in condensed matter physics.

\subsection{Barren Plateaus: The Greatest Adversary of VQE}

The barren plateau phenomenon arguably represents the most acute mathematical and operational bottleneck for the scalability of VQE. In optimization landscapes associated with highly expressive ansätze, the variance of the cost function gradient decays exponentially with the number of qubits $N$, taking the form $\text{Var}[\partial_{\theta} E] \propto \mathcal{O}(b^{-N})$ \cite{McClean2018_Barren}. This phenomenon results in a cost surface that is exponentially flat over almost its entire extent, as illustrated in Fig.~\ref{fig:desafios}(a), where the variance decay becomes intractable for generic circuits such as the Hardware Efficient Ansatz (HEA), in comparison to structured approaches, as shown in Fig.~\ref{fig:desafios}(c).

In this scenario, the gradient vanishes into the magnitude of the shot noise, and the classical optimizer becomes unable to discern the direction of the global minimum. For condensed matter simulations, this problem takes on dramatic proportions: capturing non-local correlations requires deep ansätze which, ironically, guarantee the occurrence of barren plateaus, imposing an apparent paradox of expressibility versus trainability \cite{Holmes2022_Expressibility}.

\subsection{Gradient-Based Optimization vs. Heuristic Methods}

The topology of the cost landscape in complex condensed matter systems exposes an intrinsic weakness of purely gradient-based methods. Even when barren plateaus are partially mitigated in shallow circuits, classical optimizers such as Adam or L-BFGS frequently fail to navigate geometrically rugged parameter spaces, which are riddled with spurious local minima (traps) associated with low-energy excitations of the physical system \cite{Bittel2021_NPHard}, as illustrated in Fig.~\ref{fig:desafios}(b).

From this perspective, we argue that heuristic and adaptive methods, notably the FALQON algorithm (herein denoted under the broader rubric of FQA, Feedback-based Quantum Algorithms) \cite{Magann2021, Larsen2024, Long2025_ITEFALQON}, present a substantially more robust search architecture. Instead of pre-defining a global circuit and attempting to optimize all variables simultaneously (risking vanishing gradients), FQA constructs the state evolution layer by layer in a quasi-deterministic manner. Navigation through the Hilbert space is guided by a rigorous local heuristic: the experimental measurement of the commutator between the problem Hamiltonian and the driving Hamiltonian at each time step. 

By replacing blind, global optimization in a high-dimensional space with sequential steps based on physical feedback, FQA circumvents the trap of intractable optimizations. This heuristic approach proves to be significantly more efficient for navigating regimes of quantum criticality, where fluctuations and gap closures typically disorient strict variational methods.

\subsection{Experimental Noise and the Reality of NISQ Hardware}

Finally, the transition of these algorithmic architectures to actual quantum systems introduces the issue of experimental noise, such as logic gate errors, thermal relaxation ($T_1$), dephasing ($T_2$), and readout errors, which affect each method in a distinct manner.

For VQE and QAOA, noise not only corrupts the fidelity of the prepared state but fundamentally alters the optimization landscape itself. Decoherence induces so-called noise-induced barren plateaus \cite{Wang2021_NoiseBP}, flattening the authentic global minimum and asymptotically shifting the position of the optimal parameters towards maximally mixed states, as illustrated in Fig.~\ref{fig:desafios}(d).

There is a fundamental trade-off between circuit depth and fidelity: while deeper circuits could better describe the target state, the accumulation of errors degrades fidelity exponentially in the NISQ regime, as shown in Fig.~\ref{fig:desafios}(e). Although Quantum Error Mitigation techniques, such as Zero-Noise Extrapolation and Probabilistic Error Cancellation \cite{Cai2023_QEM_Review}, are applicable, they impose a sampling overhead that is frequently prohibitive for relevant system sizes, especially when the required number of measurements (shots) already explodes due to barren plateaus, as illustrated in Fig.~\ref{fig:desafios}(f).

\section{Conclusions and Future Perspectives}

The simulation of quantum many-body systems and the characterization of Quantum Phase Transitions represent one of the most rigorous benchmarks for contemporary quantum computing. As discussed throughout this perspective, classical intractability—manifested in the dimensional explosion of the Hilbert space, the fermionic sign problem in Monte Carlo simulations, and the limitations of tensor networks in higher dimensions—positions quantum simulators as the natural platform for unraveling the physics of strongly correlated materials.

In the current NISQ era, variational algorithms such as VQE and QAOA have established the fundamental paradigm of hybrid quantum-classical optimization. However, the naive application of these methods to condensed matter problems encounters hard boundaries: the need for deep circuits to capture long-range entanglement collides head-on with hardware decoherence rates and the intractable barren plateau phenomenon. We have demonstrated that the energy landscape topology in Topological Phases, Many-Body Localization, and Quantum Spin Liquids is exceedingly complex for gradient-based optimizers to navigate efficiently, frequently trapping the simulation in spurious local minima.

In light of this scenario, we argue that the immediate future of quantum simulation in condensed matter requires a departure from blind, global optimization routines. Feedback-based Quantum Algorithms, such as FALQON, offer a powerful heuristic avenue. By replacing stochastic search with a deterministic dynamical evolution, guided layer-by-layer by local commutator measurements, FQAs mitigate the vanishing gradient problem and respect the underlying symmetries of the Hamiltonian. Nevertheless, this algorithmic resilience exacts a toll in the form of extreme sensitivity to readout noise and feedback-loop latency, defining the primary engineering challenge for their practical implementation.

Looking ahead, we envision critical fronts that will dictate the pace of discoveries at the interface between condensed matter physics and quantum information. The first resides in the co-design of algorithms and hardware. The pursuit of practical quantum advantage will require the development of physics-informed \textit{ansätze}. Instead of generic circuits, the direct mapping of gauge symmetries and exchange interactions onto the physical qubit layout will drastically reduce circuit depth and the occurrence of noise-induced barren plateaus.

Finally, the temporal trajectory of this field will be defined by the gradual transition from error mitigation to error correction. While operating in the NISQ regime, the development of more efficient Quantum Error Mitigation protocols tailored for complex many-body observables (such as dynamic structure factors and topological entanglement entropy) will be absolutely indispensable. Concurrently, the insights gained from FQAs and shallow variational algorithms will serve as crucial stepping stones during the migration towards Fault-Tolerant Quantum Computing architectures, where rigorous Hamiltonian dynamics simulation and Quantum Phase Estimation can ultimately be applied to large-scale systems.

In summary, the limitations of the NISQ era should not be viewed as a dead end, but rather as a rigorous filter that compels the physics community to develop more elegant and robust simulation methods. Overcoming barren plateaus and taming experimental noise through adaptive heuristic approaches will not only render quantum computers viable as analytical tools but will inevitably reveal new and profound facets of entanglement and criticality in many-body physics.

%
%

\ack{
G. E. L. P. acknowledges support from the Fundação de Amparo à Pesquisa do Estado de São Paulo (FAPESP), grant no.~2025/22498-8 (fellowship associated with grant no.~2024/08433-8). 
L. A. M. R. acknowledges support from FAPESP, grant no.~2025/25024-7. 
P. M. P. acknowledges support from FAPESP, grant no.~2023/12110-7. 
A. R. F. acknowledges support from FAPESP, grant no.~2024/21658-9 (fellowship associated with grant no.~2024/08433-8). 
F. F. F. acknowledges support from FAPESP, grant nos.~2024/00998-6 (CRISQuaM -- FAPESP Research, Innovation and Dissemination Center, RIDC/CEPID) and 2025/15490-0 (QuantaNet -- FAPESP Thematic Project Grant). F. F. F. also acknowledges partial financial support from the Conselho Nacional de Desenvolvimento Científico e Tecnológico (CNPq) through the National Institute of Science and Technology for Applied Quantum Computing, Grant No.~408884/2024-0, and from the Office of Naval Research (ONR), Project No.~N62909-24-1-2012.
}

\funding{
This work was funded by the São Paulo Research Foundation (FAPESP) through Grant Nos.~2025/22498-8 (G. E. L. P., fellowship associated with grant no.~2024/08433-8), 2025/25024-7 (L. A. M. R.), 2023/12110-7 (P. M. P.), 2024/21658-9 (A. R. F., fellowship associated with grant no.~2024/08433-8), 2024/00998-6 (F. F. F.; CRISQuaM -- FAPESP Research, Innovation and Dissemination Center, RIDC/CEPID), and 2025/15490-0 (F. F. F.; QuantaNet -- FAPESP Thematic Project Grant). F. F. F. also acknowledges partial financial support from the Conselho Nacional de Desenvolvimento Científico e Tecnológico (CNPq) through the National Institute of Science and Technology for Applied Quantum Computing, Grant No.~408884/2024-0, and from the Office of Naval Research (ONR), Project No.~N62909-24-1-2012.
}

\data{No new data were created or analysed in this study.}

\bibliographystyle{iopart-num} 


\bibliography{refs}

\end{document}